\documentclass[11pt]{article}
\usepackage{latexsym}
\usepackage{amsmath}\usepackage{amssymb}
\usepackage[linesnumbered,lined,algoruled,noresetcount]{algorithm2e}

\usepackage{gefpaper}

\newcommand{\set}[1]{\{{#1}\}}
\newcommand{\abs}[1]{|{#1}|}

\newcommand{\qeq}{\stackrel{{\scriptscriptstyle\rm ?}}{=}}

\newcommand{\qi}{q_{\scriptscriptstyle\rm I}}
\newcommand{\qf}{q_{\scriptscriptstyle\rm F}}

\newcommand{\gap}{\ensuremath{\mathrm{GAP}}}
\newcommand{\agap}{\ensuremath{\mathrm{AGAP}}}

\newcommand{\dgap}{\ensuremath{D_{\gap}}}
\newcommand{\ugap}{\ensuremath{U_{\gap}}}
\newcommand{\dagap}{\ensuremath{D_{\agap}}}
\newcommand{\nagap}{\ensuremath{N_{\agap}}}

\newcommand{\logspace}{\ensuremath{\mathrm{L}}}
\newcommand{\nlogspace}{\ensuremath{\mathrm{NL}}}

\newcommand{\ptime}{\ensuremath{\mathrm{P}}}

\newcommand{\tw}{\textsc{2}}

\newcommand{\twdfa}{\tw\textsc{dfa}}
\newcommand{\twdfas}{\tw\textsc{dfa}s}

\newcommand{\twufa}{\tw\textsc{ufa}}

\newcommand{\nfas}{\textsc{nfa}s}
\newcommand{\twnfa}{\tw\textsc{nfa}}
\newcommand{\twnfas}{\tw\textsc{nfa}s}

\newcommand{\twsvfa}{\tw\textsc{svfa}}
\newcommand{\twsvfas}{\tw\textsc{svfa}s}

\newcommand{\twafa}{\tw\textsc{afa}}
\newcommand{\twafas}{\tw\textsc{afa}s}

\newcommand{\twonfa}{\tw\textsc{onfa}}
\newcommand{\twonfas}{\tw\textsc{onfa}s}
\newcommand{\twoafa}{\tw\textsc{oafa}}
\newcommand{\twoafas}{\tw\textsc{oafa}s}

\newcommand{\reach}{\mbox{\sc Reach}}
\newcommand{\treach}{\mbox{\rm t\sc Reach}}
\newcommand{\nreach}{\mbox{\rm n\sc Reach}}
\newcommand{\reachable}{\mbox{\sc Reachable}}

\newcommand{\qyes}{\ensuremath{q_{\rm yes}}}
\newcommand{\qno}{\ensuremath{q_{\rm no}}}

\newcommand{\qprev}{\ensuremath{q_{\rm prev}}}
\newcommand{\qques}{\ensuremath{q_{\scriptscriptstyle\rm ?}}}
\newcommand{\Ql}[1]{\ensuremath{#1{\scriptstyle\nwarrow}}}
\newcommand{\Qr}[1]{\ensuremath{#1{\scriptstyle\nearrow}}}
\newcommand{\Qhl}[1]{\ensuremath{#1{\scriptstyle{\downarrow_1}}}}
\newcommand{\Qhr}[1]{\ensuremath{#1{\scriptstyle{\downarrow_2}}}}
\newcommand{\ql}{\Ql{q}}
\newcommand{\qr}{\Qr{q}}
\newcommand{\qhl}{\Qhl{q}}
\newcommand{\qhr}{\Qhr{q}}

\newcommand{\bsection}[2]{\section[#1]{\hspace{-1em}.\hspace{0.5em}#1\label{#2}}\ignorespaces}

\newtheorem{theorem}{Theorem}[section]
\newcommand{\btheorem}[1]{\begin{theorem}\label{#1}\ignorespaces}
\newcommand{\bTheorem}[2]{\begin{theorem}[#2]\label{#1}\ignorespaces}
\newcommand{\etheorem}{\end{theorem}}

\newtheorem{lemma}[theorem]{Lemma}
\newcommand{\blemma}[1]{\begin{lemma}\label{#1}\ignorespaces}
\newcommand{\elemma}{\end{lemma}}

\newtheorem{corollary}[theorem]{Corollary}
\newcommand{\bcorollary}[1]{\begin{corollary}\label{#1}\ignorespaces}
\newcommand{\ecorollary}{\end{corollary}}

\newcommand*{\benumerate}{\begin{enumerate}}
\newcommand*{\eenumerate}{\end{enumerate}}

\renewcommand*{\theenumi}{{\roman{enumi}}}

\newcommand*{\bitemize}{\begin{itemize}}
\newcommand*{\eitemize}{\end{itemize}}

\newenvironment{proof}{\noindent\emph{Proof\/}:\hspace{\labelsep}\ignorespaces}{\bigbreak}
\newcommand*{\qed}{\mbox{}\nolinebreak\hfill~\raisebox{0.77ex}[0ex]{\framebox[1ex][l]{}}}
\newcommand*{\bproof}{\begin{proof}}
\newcommand*{\eproof}{\qed\end{proof}}

\SetVlineSkip{0pt}
\SetAlgoVlined

\begin{document}
\title{\mbox{}\\[-1.5cm]%
Two-Way Automata Making Choices Only at the Endmarkers%
\footnotetext[1]{Supported by the Slovak Grant Agen\-cy for
  Science (VEGA) under contract ``Combinatorial
  Structures and Complexity of Algorithms''$\!$,
  and by the Slovak Research and Development Agency (APVV) under contract
  ``Algorithms, Automata, and Discrete Data Structures''$\!$\@.}%
}%
\author{Viliam Geffert\footnotemark[1]\and
Bruno Guillon\footnotemark[2]\and
Giovanni Pighizzini\footnotemark[3]\and
  \mbox{}\\
{\normalsize \mbox{\footnotemark[1] }\,Department of Computer
    Science -- P.\,J.~\v{S}af\'{a}rik University, Ko\v{s}ice -- Slovakia}\\
  {\small\sf viliam.geffert@upjs.sk}\\[1.0ex]%
{\normalsize \mbox{\footnotemark[2] }\,Universit\'e Nice-Sophia Antipolis and
   \'Ecole Normale Sup\'erieure de Lyon -- France}\\
  {\small\sf guillon.bruno+cs@gmail.com}\\[1.0ex]%
{\normalsize \mbox{\footnotemark[3] }\,Dipartimento di
    Informatica e Comunicazione -- Universit\`{a} degli Studi di
    Milano -- Italy}\\
  {\small\sf pighizzini@dico.unimi.it}%
}%
\date{}%
\maketitle\thispagestyle{empty}
\begin{quotation}\small\noindent
  \textbf{Abstract.}\hspace{\labelsep}%
The question of the state-size cost for simulation of two-way
nondeterministic automata (\twnfas) by two-way deterministic
automata (\twdfas) was raised in 1978 and, despite many attempts,
it is still open. Subsequently, the problem was attacked by
restricting the power of \twdfas\ (e.g., using a restricted input
head movement) to the degree for which it was already possible to
derive some exponential gaps between the weaker model and the
standard \twnfas\@. Here we use an opposite approach, increasing
the power of \twdfas\ to the degree for which it is still
possible to obtain a subexponential conversion {}from the
stronger model to the standard \twdfas\@. In particular, it turns
out that subexponential conversion is possible for two-way
automata that make nondeterministic choices only when the input
head scans one of the input tape endmarkers. However, there is no
restriction on the input head movement. This implies that an
exponential gap between \twnfas\ and \twdfas\ can be obtained
only for unrestricted \twnfas\ using capabilities beyond the
proposed new model.

As an additional bonus, conversion into a machine for the
complement of the original language is polynomial in this model.
The same holds for making such machines self-verifying, halting,
or unambiguous. Finally, any superpolynomial lower bound for the
simulation of such machines by standard \twdfas\ would imply
$\logspace\neq\nlogspace$\@. In the same way, the alternating
version of these machines is related to
$\logspace\qeq\nlogspace\qeq\ptime$, the classical computational
complexity problems.\\
\mbox{}\\
\textbf{Keywords:}\hspace{\labelsep}%
two-way automata, descriptional complexity, regular languages
\end{quotation}

\bsection{Introduction}{s:intr}
The cost, in terms of states, of the simulation of two-way
nondeterministic automata (\twnfas, for short) by two-way
deterministic automata (\twdfas) is one of the most important and
challenging open problems in automata theory and, in general, in
theoretical computer science. This problem was proposed in~1978
by Sakoda and Sipser~\cite{SS78}, who conjectured that the cost
is exponential. However, in spite of all effort, exponential gaps
were proved only between \twnfas\ and some restricted
\emph{weaker versions of \twdfas}\@.

In~1980, Sipser proved that if the resulting machine is required
to be \emph{sweeping} (deterministic and reversing the direction
of its input head only at the \emph{endmarkers}, two special
symbols used to mark the left and right ends of the input), the
simulation of a \twnfa\ is indeed exponential~\cite{Sip80}\@.
However, Berman and Micali~\cite{Ber80,Mic81} proved
independently that this does not solve the general problem: in
fact the simulation of unrestricted \twdfas\ by sweeping \twdfas\
also requires an exponential number of states. The Sipser's
result was generalized by Hromkovi\v{c} and
Schnitger~\cite{HS03}, who considered \emph{oblivious machines}
(following the same trajectory of input head movements along all
inputs of equal length) and, recently, by
Kapoutsis~\cite{Kap11b}, considering \twdfas\ with the number of
input head reversals that is sublinear in the length of the
input. However, even the last condition gives a machine provably
less succinct than unrestricted \twdfas, and hence the general
problem remains open.

\medskip
Starting {}from~2003 with a paper by Geffert
\emph{et~al.}~\cite{GMP03}, a different kind of restriction has
been investigated: the subclass or regular languages using
\emph{a single-letter input alphabet}\@. Even under this
restriction, the problem of Sakoda and Sipser looks difficult,
since it is connected with $\logspace\qeq\nlogspace$, an open
question in complexity theory. (\logspace~and \nlogspace\ denote
the respective classes of languages accepted in deterministic and
nondeterministic logarithmic space\@.) First, in~\cite{GMP03}, a
new normal form was obtained for unary automata, in which all
nondeterministic choices and input head reversals take place only
at the endmarkers. Moreover, the state-size cost of the
conversion into this normal form is only \emph{linear}\@. This
normal form is a starting point for several other properties of
unary \twnfas\@. First, in the same paper, each $n$-state unary
\twnfa\ is simulated by an equivalent \twdfa\ with
$O(n^{\lceil\log_2(n+1)+3\rceil})$ states, which gives a
subexponential but still superpolynomial upper bound. It is not
known whether this simulation is tight. However, a positive
answer would imply the separation between the classes \logspace\
and \nlogspace\@. In fact, under assumption that
$\logspace=\nlogspace$, each unary \twnfa\ with $n$ states can be
simulated by a \twdfa\ with a number of states polynomial
in~$n$~\cite{GP11}\@. After a minor modification (without
assuming $\logspace=\nlogspace$), this gives that each unary
\twnfa\ can be made unambiguous, keeping the number of the states
polynomial. (For further connections between two-way automata and
logarithmic space, we address the reader
to~\cite{BL77,Kap11a}\@.)

Along these lines of investigation, in~\cite{GMP07}, the problem
of the complementation for unary \twnfas\ has been considered, by
proving that each $n$-state \twnfa\ accepting a unary
language~$L$ can be replaced by a \twnfa\ with $O(n^8)$ states
accepting the complement of~$L$\@. The proof combines the above
normal form for unary \twnfas\ with inductive counting arguments.

Kapoutsis~\cite{Kap06} considered the complementation in the case
general input alphabets, but restricting the input head
reversals. He showed that the complementation of \emph{sweeping}
\twnfas\ (with the input head reversals only at the endmarkers)
requires exponentially many states, thus emphasizing a relevant
difference with the unary case.

\medbreak
In this paper, we use a different approach. Instead of
\emph{restricting the power of \twdfas} to the degree for which
it is already possible to derive an exponential gap between the
weaker model and the standard \twnfas, we \emph{increase the
power of \twdfas}, towards \twnfas, to the degree for which it is
still possible to obtain a subexponential conversion {}from the
stronger model to the standard \twdfas\@. Such new stronger model
then clearly shows that, in order to prove an exponential gap
between \twnfas\ and \twdfas, one must use capabilities not
allowed in the proposed new model. More precisely, in our new
model,  we neither
restrict the cardinality of input alphabets, nor put any
constraint on the head movement, i.e., head reversals can take
place at any input position. On the other hand, we permit
nondeterministic choices only when the input head is scanning one
of the endmarkers. We shall call such machine a \emph{two-way
outer-nondeterministic finite automaton} (\twonfa)\@.

It turns out that this machine has its natural counterpart also
in the case of \emph{two-way alternating finite automata}
(\twafas), which is a \emph{two-way outer-alternating finite
automaton} (\twoafa), making universal and existential choices
only at the endmarkers. (For recent results on \twafas,
see~\cite{Kap09,Gef11}\@.)

We show that several results obtained for \emph{unary} \twnfas\
can be extended to \twonfas, and some of them even to the
alternating version, \twoafas, with \emph{any} input alphabet. In
particular, we prove the following:
\bitemize
  \item Each $n$-state \twonfa\ can be simulated by a
    \emph{halting} two-way \emph{self-verifying}
    automaton~(\twsvfa)~\cite{DHRS97} with $O(n^8)$ many states.
    This fact has two important implications:
    \bitemize
      \item The complementation of \twonfas\ can be done by using
        a polynomial number of states. Note the contrast with the above
        mentioned case of sweeping \twnfas\ studied in~\cite{Kap06}\@.
      \item Each \twonfa\ can be simulated by a \emph{halting} \twonfa\
        using a polynomial number of states.
    \eitemize
  \item Each $n$-state \twonfa\ can be simulated by a \twdfa\ with
    $O(n^{\log_2n+6})$ states.
  \item If $\logspace=\nlogspace$, then each $n$-state \twonfa\ can
    be simulated by a \twdfa\ with a number of states polynomial
    in~$n$\@. Hence, a superpolynomial lower bound for the simulation
    of \twonfas\ by \twdfas\ would imply $\logspace\neq\nlogspace$\@.
    (Unlike in~\cite{BL77}, there are no restrictions on the length of
    potential witness inputs\@.)
  \item Each $n$-state \twonfa\ can be simulated by an
     \emph{unambiguous} \twonfa\ with a polynomial number of states.
  \item If $\logspace=\ptime$, then each $n$-state \twoafa\ can be
    simulated by a \twdfa\ with a number of states polynomial in~$n$,
    with the same consequences as presented for
    $\logspace\qeq\nlogspace$\@. (\ptime~denotes, as usual, the class
    of languages recognizable by deterministic Turing machines in
    polynomial time\@.)
  \item Similarly, if $\nlogspace=\ptime$, we get the
    corresponding polynomial conversion {}from \twoafas\ to \twnfas\@.
\eitemize

These results are obtained by generalizing the constructions
given in~\cite{GMP03,GMP07,GP11} for the unary case. However,
here we do not have a normal form for simplifying the automata,
by restricting input head reversals to the endmarkers. Our
generalization rely on a different tool, presented in the first
part of the paper. Basically, we extend some techniques developed
originally for deterministic devices~\cite{Sip80h,GMP07} to
machines with nondeterminism at the endmarkers. This permits us
to check the existence of certain computation paths, including
infinite loops, by the use of a linear number of states.

\medskip
The paper is organized as follows. In Section~\ref{s:prel} we
recall basic definitions and preliminary results required later.
In Section~\ref{s:reach}, after introducing \twonfas, we develop
a fundamental tool that will be used several times, namely, a
deterministic procedure that allows us to check the existence of
computation paths between two given states in the given \twonfa,
starting and ending at the left endmarker and do not visiting the
endmarkers in the meantime. This procedure is also useful to make
all computations halting. The next sections are devoted to our
main results. In Section~\ref{s:compl}, we present the polynomial
simulation of \twonfas\ by \twsvfas\ and its consequences. In
Section~\ref{s:deter}, we show the subexponential simulation of
\twonfas\ by \twdfas. Then, in Section~\ref{s:gap}, under the
assumption $\logspace=\nlogspace$, such simulation is with a
polynomial number of states. Furthermore, we show how to simulate
a \twonfa\ by an unambiguous \twnfa\ using a polynomial number of
states. Finally, in Section~\ref{s:alt}, we present the
corresponding results for the alternating case. Some concluding
remarks are briefly discussed in Section~\ref{s:conclusion}\@.

\bsection{Preliminaries}{s:prel}
Let us start by briefly recalling some basic definitions {}from
automata theory. For a detailed exposition, we refer the reader
to~\cite{HU79}\@. Given a set~$S$, $\abs{S}$~denotes its
cardinality and $2^S$ the family of all its subsets.

A \emph{two-way nondeterministic finite automaton} (\twnfa, for
short) is defined as a quintuple $\mathcal{A}=(Q,\Sigma,\delta,\qi,F)$, in
which $Q$ is a finite set of states, $\Sigma$~is a finite input
alphabet,
$\delta:Q\times(\Sigma\cup\set{\vdash,\dashv})\rightarrow
2^{Q\times\set{-1,0,+1}}$
is a transition function, where $\vdash,\dashv\;\notin\Sigma$ are
two special symbols, called the left and the right endmarkers,
respectively, $\qi\in Q$ is an initial state, and $F\subseteq Q$
is a set of final states. The input is stored onto the input tape
surrounded by the two endmarkers, the left endmarker being at the
position zero. In one move, $\mathcal{A}$~reads an input symbol, changes
its state, and moves the input head one position forward,
backward, or keeps it stationary depending on whether $\delta$
returns $+1$, $-1$, or~$0$, respectively. The machine accepts the
input, if there exists a computation path {}from the initial
state~$\qi$ with the head on the left endmarker to some final
state $q\in F$\@. The language accepted by~$\mathcal{A}$, denoted
by~$L(\mathcal{A})$, consists of all input strings that are accepted.
$\mathcal{A}$ is said to be \emph{halting} if each computation ends
in a finite number of steps.

Observing that if an accepting computation visits the same endmarker
two times in the same state then there exists a shorter accepting
computation on same input, we immediately get the
following lemma, which will be used in the proofs of some of our results:

\blemma{l:bound}
  If a \twnfa\ $\mathcal{A}$ with $n$ states accepts an input
  $w$, then it also accept $w$ with a computation
  that visits the left (right) endmarker at most $n$ times.
\elemma

Throughout the paper, given a \twnfa\ $\mathcal{A}$,
we will call \emph{computation segment} (or just segment)
{}from $p$ to $q$ on $w$, each computation path on an input $w$
that starts at the left endmarker in the state $p$,
ends at the left endmarker in the state $q$ and never visits
the same endmarker in the meantime.

A \emph{sequence of $t\geq 0$ segments} {}from
a state $p$ to a state $q$ on input $w$ is a sequence of
segments such that there are states
$p_0,p_1,\ldots,p_t$, with $p_0=p$, $p_t=q$, and
the $i$th segment is {}from $p_{i-1}$ to $p_i$,
$i=1,\ldots,t$.

A \twnfa\ $\mathcal{A}$~is said to be \emph{deterministic} (\twdfa),
whenever $\abs{\delta(q,\sigma)}\le 1$, for any $q\in Q$ and
$\sigma\in\Sigma\cup\set{\vdash,\dashv}$\@,~$\mathcal{A}$ is called
\emph{unambiguous} (\twufa), if there exists at most one accepting
computation path for each input.
A two-way \emph{self-verifying} automaton (\twsvfa) $\mathcal{A}$ is a
\twnfa\ which, besides the set of accepting states $F\subseteq Q$, is
equipped also with a 
disjoint
set of rejecting states $F^r\subseteq Q$.
For each input $w\in L(\mathcal{A})$ there exists one computation path
halting in a state $q\in F$, and no path may halt in a state
$q\in F^r$. Conversely, for $w\notin L(\mathcal{A})$ there exists one computation
path halting in a state $q\in F^r$, and no path may halt in
a state $q\in F$. Note that some computation paths of a
\twsvfa\ may end with a ``don't-know'' answer, by ending in
one state not belonging to $F\cup F^r$, or entering into
an infinite loop.

An automaton working over a single letter alphabet is called
\emph{unary}\@.

\medbreak
We assume that the reader is familiar with the standard Turing
machine model and the basic facts {}from space complexity theory.
For more details, see e.g.~\cite{HU79,Sze94}\@. Our Turing
machine (both deterministic and nondeterministic) is equipped
with a finite state control, a two-way read-only input tape, and a separate
semi-infinite two-way read-write worktape, initially empty. Such
machine is \emph{$s(n)$-space bounded}, if no computation uses
more than $s(n)$ worktape cells, for each input of length~$n$\@.

The class of languages accepted in $s(n)\le O(\log n)$ space by
deterministic Turing machines is denoted by \logspace, while the
corresponding classes for nondeterministic
machines by~\nlogspace\@.

\bsection{Outer-nondeterministic automata and the subroutine \reach}{s:reach}

In this section we introduce the computational model investigated in the
paper, and we develop some preliminary results concerning it. First,
we provide a simplification of the model with respect to the definition.
After that, we will present the main result of this section.
We will show how to detect the existence of computational segments in our devices.
This result will be a fundamental tool, used in the proof of all
the other results presented in the paper.

A \emph{two-way outer-nondeterministic finite automaton}
(\twonfa, for short) is a \twnfa~$\mathcal{A}=(Q,\Sigma,\delta,\qi,F)$
that can take nondeterministic
decisions only when the input head is scanning one of the two
endmarkers, i.e., for each $q\in Q$, $a\in\Sigma$,
$\abs{\delta(q,a)}\leq 1$.
Actually, with a linear increasing in the number of the states,
we can further restrict the use of the nondeterminism to the
left endmarker only. We can also obtain some other restrictions,
which will be useful to  simplify our proofs:

\blemma{l:normalform}
  For any $n$-state \twonfa~$\mathcal{A}=(Q,\Sigma,\delta,\qi,F)$
  there exists an equivalent \twonfa~$\mathcal{A}'$ with no more than $3n$ states
  that satisfies the following properties:
  \begin{enumerate}
    \item nondeterministic choices are taken only when the input head is scanning the left endmarker,
    \item there is a unique accepting state $\qf$ and this state is also halting,
    \item $\qf$ is reachable only at the left endmarker by stationary moves,
    \item stationary moves can occur only at the left endmarker to enter the state $\qf$.
  \end{enumerate}
\elemma
\bproof
  If $F=\emptyset$ then $\mathcal{A}$ accept the empty language and it can
  be replaced by the trivial automaton with two states $\qi$ and $\qf$,
  and without any transition.
  So, {}from now on let us suppose $F\neq\emptyset$.

  First, we can make each final state also halting, by making the transition function
  undefined {}from final states. Furthermore, each stationary move leading to a final state
  can be replaced by a transition to the same final state that moves the head either to the
  left or to the right.
  Now, by inspecting the transition function, for each $q\in Q$ and $a\in\Sigma\cup\set{\vdash,\dashv}$,
  we compute the set of pairs $(p,d)$ such that $\mathcal{A}$ in the state $q$ scanning an input
  square containing the symbol $a$, after a possible sequence of stationary moves,
  will finally reach the state $p$ moving the head in the direction $d\in\set{-1,1}$.
  Stationary moves {}from the state $q$ with input $a$ are
  replaced by all possible transitions described by those pairs $(p,d)$. Notice that
  if $a$ is an ordinary input symbol, i.e., $a\notin\set{\vdash,\dashv}$, then we can have at
  most one such a pair.
  Hence, after this transformation, nondeterministic choices remain confined at the endmarkers
  and the resulting automaton does not perform any stationary move.

  Now, {}from the automaton $\mathcal{A}$ so modified, we build an automatom $\mathcal{A}'$
  making nondeterministic choices only on the left endmarker.
  Normally, $\mathcal{A}'$ makes the same moves as~$\mathcal{A}$. However, when the head
  reaches the right endmarker in a state $q$, $\mathcal{A}'$ traverses the input {}from right to left,
  using a copy $\overleftarrow{q}$ of $q$, to reach the left endmarker.
  At this position, the automaton simulates a transition {}from $q$ at the right endmarker,
  i.e., it chooses a state $p$ such that $(p,-1)\in\delta(q,\dashv)$ and, using a copy $\overrightarrow{p}$
  of $p$, it traverses the input {}from left to right until to reach the right endmarker. Here,
  $\mathcal{A'}$ moves its head one position to the left entering the original state $p$.
  {}From this configuration $\mathcal{A}'$ resumes the simulations of the ordinary moves
  of~$\mathcal{A}$.

  Finally, to accept in $\qf$ on the left endmarker, we make two further small changes.
  First, when $\mathcal{A}'$ enters a state $p\in F$, it starts to move its head to
  the left using the state $\overleftarrow{p}$, until to reach the left endmarker.
  At that position, {}from $\overleftarrow{p}$, $\mathcal{A}'$ finally enters $\qf$
  without moving the input head.
  Furthermore, when~$\mathcal{A}'$, scanning the left endmarker, should move in
  a state $\overrightarrow{p}$, for some $p\in F$, then it enters directly $\qf$ without moving the
  input head. This second change permit us to save the states $\overrightarrow{p}$, for $p\in F$.

  It can be easily seen that the resulting automaton $\mathcal{A}'$ is equivalent to the given
  \twonfa~$\mathcal{A}$ and satisfies the properties listed in the statement
  of the lemma.
  The set of states consists of $3$ copies of the set $Q$ plus the state $\qf$. However, for each final
  state $p$ of $\mathcal{A}'$, the copy $\overrightarrow{p}$ is useless.
  Hence, the total number of states of $\mathcal{A}'$ does not exceed $3n$.
\eproof

The remaining part of this section is devoted to develop a tool which will be fundamental
in the proof of our results.
Given a \twonfa\ $\mathcal{A}$ with $n$ states, we show the existence of a subroutine \reach\
that receives as parameters two states $q',q''$ of $\mathcal{A}$ and decides whether or not
$\mathcal{A}$ has a computation segment {}from $q'$ to $q''$ on an input string $w$.
We will show that this subroutine can be implemented using a deterministic finite state
control with $O(n)$ states whose input $w$ is stored on a two-way tape.

At a first glance, we could try to compute $\reach(q',q'')$ by initializing the
automaton~$\mathcal{A}$ in the state
$q'$ with the input head at the left endmarker and by stopping its computation
as soon as the input head, in one of the following steps, reaches again the left endmarker, then testing
whether or not the state so reached is $q''$.
However, this approach presents two problems: first of all, the original automaton $\mathcal{A}$
could enter into a infinite loop, never coming back to the left endmarker;
second, the first move {}from the state $q'$ on the left
endmarker can be nondeterministic.

In order to solve the first problem, we adapt the construction given in~\cite{GMP07} to transform each $n$-state
\twdfa\ into an equivalent halting $4n$-state \twdfa, which, it turns, was a refinement of the construction
obtained by Sipser~\cite{Sip80h} to make space bounded Turing machines halting. We give a brief outline.
For each $w\in\Sigma^*$,  a deterministic machine accepts $w$ if and only if there is a ``backward'' path,
following the history of the computation in reverse, {}from the unique accepting configuration $c_f$ to the
unique initial configuration $c_0$. In our setting, a ``configuration'' is a pair $(s,i)$, where $s\in Q$ is
a state and $i\in\set{0,\ldots,|w|+1}$ is an input head position.

Consider the graph whose nodes represent configurations and edges computation steps.
If the machine under consideration is deterministic, the component of the graph containing the accepting configuration
$c_f$ is a tree rooted at this configuration, with backward paths branching to all possible predecessors of $c_f$.
In addition, if the accepting configuration is also halting, no backward path starting {}from
$c_f$ can cycle (hence, it is of finite length).
Thus, the equivalent halting machine can perform a depth-first search of this tree
in order to detect whether the initial configuration $c_0$ belongs to the predecessors of $c_f$.
If this is the case, then the simulator accepts.
On the other hand, if all the tree is examined without reaching $c_0$, this means that there are no paths {}from
$c_0$ to $c_f$ and so $w$ is not in the language. Hence, the simulator rejects.

We adapt such a procedure by choosing $c_0=(q',0)$ and $c_f=(q'',0)$, where $q'$ and $q''$ are the
two parameters. Furthermore, since we are interested to detect the existence of \emph{just one
segment} from $q'$ to $q''$, we do not consider the transitions on the left endmarker
{}from states different than $q'$. However, the remaining transition, 
i.e., that {}from configuration $c_0$, can be nondeterministic.
Thus, we need further modifications.
Suppose $\delta(q',\vdash)=\set{(q_1,+1),\ldots,(q_k,+1)}$. Hence, there is a segment {}from
$c_0$ to $c_f$ if and only if there is a path {}from $c_j=(q_j,1)$ to $c_f$, for some $j\in\{1,\ldots,k\}$,
visiting the left endmarker only in $c_f$.
Hence, the backward search can be done starting {}from $c_f$, without considering all the transitions
of the original automaton {}from the left endmarker, stopping and accepting when one of the
$c_j$'s is reached, or rejecting when all the tree has been visited without reaching any of the $c_j$'s.

To do that, our procedure needs to detect when the head position of the original
\twonfa\ $\mathcal{A}$ is scanning the first ``real'' input symbol, i.e., that immediately
to the right of the left endmarker, in cell number $1$. This can be done as follows.
We say that a \emph{left predecessor} for the configuration $(s,i)$ is a configuration $(t,i-1)$,
with the input head scanning a symbol $a$, such that $(s,+1)\in\delta(t,a)$.
Hence, in the construction of the backward path {}from $(q'',0)$ to $(q',0)$, the configuration
$(q',0)$ can be reached only as a left predecessor of some $(q_j,1)$, $j\in\set{1,\ldots,k}$.
By a closer look to the simulation in~\cite{GMP07}, we can observe that all the left precedessors of
a configuration $(s,i)$ are examined in a copy $\Ql{s}$ of the state $s$, with the input head
one position to the left of the actual position of the original machine, i.e., in position $i-1$.
Hence, when the simulating machine reaches one of the states $\Ql{q_1},\ldots,\Ql{q_k}$, with the
input head scanning the left endmarker at the same time, it can stop the computation and accept.
There are only two points which depend on the states $q',q''$: the choice of the initial state,
i.e., of the configuration $c_f$, and the detection of configurations $c_j=(q_j,1)$ of $\mathcal{A}$, when
the head of $\mathcal{A}'$ is scanning the left endmarker.

This construction can be implemented by a deterministic finite state control with $4n-3$ states
(not counting the space needed to store the parameters $q',q''$, which will be used in a
``read-only'' way). More details are presented in Appendix~\ref{a:reach}.
Hence, with this strategy we obtain the following result which will be a fundamental
tool in the next sections:

\blemma{l:lemma}
  For each \twonfa\ $\mathcal{A}$ with $n$ states (in the form of Lemma~\ref{l:normalform}),
  it is possible to construct a deterministic finite state control with $4n-3$ states that given
  an input string $w\in\Sigma^*$ stored on a two-way tape and two states $q',q''$
  of $\mathcal{A}$ decides whether or not the automaton
  $\mathcal{A}$ has a computation segment {}from $q'$ to $q''$ on $w$.
\elemma

\bsection{Simulation by halting self-verifying automata}{s:compl}

In this section we prove that for each $n$-state \twonfa\ $\mathcal{A}$ accepting a language $L$
there exists an equivalent \emph{halting} \twsvfa\ using a polynomial number of states and
making nondeterministic choices only when the input head is scanning the left endmarker.
As a consequence, we can derive halting \twonfas\ with polynomial many states that accept $L$
and the complement of~$L$.

Even in this case, our starting point is a proof given in~\cite{GMP07} for the unary case,
which was based on the well known technique of the \emph{inductive counting}.
However, there are deep differences. In particular, the proof in~\cite{GMP07} uses a
normal form for unary \twnfas\ in which a computation is a sequence
of deterministic traversals of the input {}from one endmarker to the opposite one.
In this normal form there are no parts of computations starting and ending at the same endmarker
without visiting the other one in the meantime (these parts are usually called \emph{U-turns}).
Furthermore, the only possible infinite loops involve the endmarkers and can be easily avoided using
Lemma~\ref{l:bound}.
The simulation inductively counts, for increasing values of $t$, how many states are reachable {}from the
initial state in $t$ traversals of the input and, as a side effect of this counting
procedure, also lists these states. In this way, after a certain number of traversals, all the
states which are reachable at the endmarkers have been listed and, hence, it is possible to decide whether or
not the input was accepted by the original machine.

In the case we are considering, we do not have such a kind of normal form. Hence, a computation can
present traversals of the input {}from one endmarker to the opposite one as well as $U$-turns.
Furthermore, a computation can reject by entering into a infinite loop: in this case infinite
loops which never visits the endmarkers are also possible. To overcome the first
problem, instead of traversals, our inductive counting procedure considers computation segments:
for increasing values of $t$, it counts how many states are reachable {}from the initial state in a computation
consisting of $t$ segments, i.e., visiting the left endmarker $t+1$ times, and, at the same time,
the procedure lists these states.
Since, as stated in Lemma~\ref{l:bound},
each accepted input has an accepting computation which visits the left endmarker at most $n$ times,
it is enough to consider computations consisting of at most $n-1$ segments. This avoid infinite loops
involving the endmarkers. We will discuss later on how to deal with the remaining infinite loops,
namely those that do not involve endmarkers.

In some steps, the simulation we present can reach the ``don't-know'' state~\qques.
We always assume that this state
is halting: hence, when it is reached, even in the code of the subroutines,
the \emph{entire} simulation will be aborted.

To simulate single computation segments, we will make use of the subroutine $\reach$, discussed in
Section~\ref{s:reach}. Furthermore we use another subroutine $\treach$ which with parameters
$q\in Q$ and $t\geq 0$ verifies the existence of sequence of $t$ segments {}from the initial
state~$\qi$ to $q$ on the input under consideration.
In negative case, $\treach$ aborts the simulation in the state~\qques,
otherwise the subroutine returns to the main simulation in a different state.
(As we will see, this subroutine is nondeterministic. It can also halt in the state~\qques, due
to a wrong sequence of nondeterministic guesses. However, if there exists a sequence of $t$ segments {}from $\qi$
to $q$, the subroutine has at least one computation ending in a state other than~\qques.)

\bigskip
\SetVlineSkip{0pt}
\SetAlgoVlined
\begin{algorithm}[H]
  \caption{\emph{Simulation of \twonfas\ by \twsvfas}}
  \small
\DontPrintSemicolon
  $m':= 1$\label{rntosv_initm'}\;
  \For{$t:= 0 \ \KwTo \ |Q|-2$\label{rntosv_tloop}}{
      $m := m'$\label{rntosv_savem'}; $m' := 0$\;
      \ForEach{$q' \in Q$\label{rntosv_qloop}}{
          \For{$i:= 1 \ \KwTo \ m$\label{rntosv_iloop}}{
            $q:=$ a nondeterministically chosen state\label{rntosv_choice}\;
            \lIf{$ (i > 1 \ and \ q \leq\qprev)$\label{rntosv_test}}{
                halt in $\qques$\label{rntosv_idontknow}\;
            }
            $\qprev:= q$\label{rntosv_saves}\;
            \If(\\\tcp*[f]{possible side effect of \treach:\ abort in $\qques$}){$\treach(q,t)$ and $\reach(\qprev,q')$\label{rntosv_reach}}{%
                \lIf{$q' = \qf$}{
                    halt in \qyes\label{rntosv_accept}\;
                }
                $m':= m' + 1$\label{rntosv_m'plusone}\;
                $break$\label{rntosv_break}\;
            }
        }\label{rntosv_endiloop}
    }\label{rntosv_endqloop}
  }\label{rntosv_endtloop}
  halt in \qno\label{rntosv_reject}\;
\end{algorithm}
\bigskip
\noindent
The algorithm proceeds by counting, for $t=0,\ldots,|Q|-2$, the number of states reachable by~$\mathcal{A}$
at the left endmarker by all computation paths starting {}from the initial configuration and consisting of exactly
$t+1$ segments (loop from line~\ref{rntosv_tloop}).
During this process, the algorithm
also generates all states reachable at the left endmarker, and hence it can correctly decide whether to accept
or reject the given input.

At the beginning of the $t$-th iteration of this loop (line~\ref{rntosv_tloop}),
a variable $m'$ contains the number of states reachable at the left endmarker
by all computation paths with exactly $t$ segments.
(In line~\ref{rntosv_initm'}, we prepare $m'=1$ for $t=0$,
the only state reachable by $0$ segments being the initial state~$\qi$.)
In line~\ref{rntosv_savem'}, we save the ``old'' value of~$m'$ in the variable~$m$,
and clear~$m'$ for counting the number of states reachable upon completing one more segment,
i.e., with exactly $t+1$ segments.

The value of $m'$ is computed in the loop from line~\ref{rntosv_qloop}:
for each state $q'\in Q$, we test whether or not it is reachable by a path with exactly $t+1$ segments.
If it is, we increment the value of~$m'$.

To decide whether $q'$ can be reached by exactly $t+1$ segments, we generate all $m$ states that are
reachable at the left endmarker by all computation paths of exactly $t$ segments and we verify if $q'$
can be reached by any of these states by a single segment.
This part is realized by the innermost loop (from line~\ref{rntosv_iloop})

The loop generates in a nondeterministic way all the sequences of $m$ states.
To avoid to consider a same set of $m$ states twice,
or to consider sequences containing repeated states, the sequences
that do not respect a preliminary fixed order on $Q$ are discarded, aborting the computation
in~\qques\ (line~\ref{rntosv_test}).
As a side effect of the call $\treach(q,t)$ (line~\ref{rntosv_reach}),
among the remaining sequences the procedure filters the only  one consisting of
the $m$ states reachable in exactly $t$ segments, discarding the remaining.
In fact, if the subroutine does not find a sequence of $t$ segments
{}from $\qi$ to $q$, then it aborts the computation in $q_?$.
(Further details about this subroutine, which is also nondeterministic, are given below.)
Hence, the only computation which ``survive'' is the one generating, according to the
fixed order, all $m$ states reachable in exactly $t$ segments.
For each $q$ among these $m$ states, the algorithm tests whether the state $q'$ under
examination is reachable {}from $q$ in one step (call $\reach(\qprev,q')$ on line~\ref{rntosv_reach};
for implementation reasons, the first parameter in the call is $\qprev$ which was initialized
in line~\ref{rntosv_saves} with the value of $q$).
If also the result of this test is positive, then the variable $m'$ is incremented.
At this point we can abort the innermost loop (line~\ref{rntosv_break}) and continue with
the next iteration, if any, of the loop from line~\ref{rntosv_qloop},
to state the reachability in $t+1$ segments of another state.
Furthermore, if during this process we discover that the final state is reachable at the left endmarker,
we stop the computation and accept by halting in the state~\qyes.

On the other hand, if the iteration of the outermost loop has been completed for each $t=0,\ldots,|Q|-2$,
never reaching $\qf$ at the left endmarker, then the input is not accepted by the original
automaton. (Otherwise, the search would have stopped already, in line~\ref{rntosv_accept}.)
Therefore, in line~\ref{rntosv_reject}, we stop in the rejecting state~\qno.

\medbreak
It is not hard to see that:
(i)~if the input is accepted by~$\mathcal{A}$, at least one computation path halts in the state~\qyes,
and no path halts in~\qno,
(ii)~if the input is rejected, at least one path halts in~\qno,
and no path halts in~\qyes\@.
(iii)~Due to wrong sequences of nondeterministic guesses,
some computation paths halt in~\qques, but no path can get into an infinite loop.

\medskip

Finally, we briefly discuss a possible implementation of the subroutine $\treach$
(for further details see Appendix~\ref{a:treach}).
Firstly, we can modify the backward seach described in Section~\ref{s:reach},
to obtain subroutine that given as parameter a state $q''$
outputs a nondeterministically chosen state $q'$ such that $\mathcal{A}$
has a computation segment {}from $q'$ to $q''$ on the given input $w$.
This subroutine, called $\nreach$, can also end in the state~\qques, aborting
the entire computation, if it is not able to find such a state $q'$.
This can  be done using $4n-4$ states, besides~\qques.
The implementation of $\treach(q,t)$ consists of a loop of $t$ iterations.
In each one of them, $\nreach$ is called on the parameter~$q$,
assigning to it the result.
Hence, at the end, if $q$ contains the initial state then the
search was successful and  $\treach$ returns the control to the caller, implicitly
giving a positive answer, otherwise it aborts the entire computation in~\qques.
Instead of this approch, we could use, as in~\cite[subroutine $simulation$]{GMP07},
a direct nondeterministic simulation of~$\mathcal{A}$, in order to generate all
the states reachable in exactly $t$
segments. This will also produce a correct inductive counting procedure.
However, due to the fact that $\mathcal{A}$ can enter some infinite loop
that does not visit the endmarkers, the resulting procedure could also enter some
loop. Our implementation produces an halting automata, obtaining,
as shown below, the same upper bound for the number of the states, as the
implementation in~\cite{GMP07} for the unary case.

The number of possible values of each one of the $6$ variables $m,m',t,q,q',\qprev$ is bounded
by $n+1$. The subroutine $\treach$ uses one counter bounded by $n$ and $O(n)$ states
to run $\nreach$, besides the ``global'' variable $q$, hence $O(n^2)$ states. The implementation of
$\reach$ uses $O(n)$ states, however, it can recycle part of the space used by $\treach$.
Hence, all the simulation can be carried out using $O(n^8)$ states.

\medskip
Finally, we  also observe that in the main algorithm and in the subroutines all the nondeterministic
choices can be taken when the input head is scanning the left endmarker.

By summarizing, we have proved the following:

\btheorem{t:svfa}
  Each $n$-state \twonfa\ can be simulated by an equivalent halting $O(n^8)$-state \twsvfa\ making
  nondeterministic choices only when the input head is scanning the left endmarker.
\etheorem

\bcorollary{c:compl}
  For each $n$-state \twonfa\ $\mathcal{A}$ there exist an equivalent halting \twonfa\ $\mathcal{A}'$
  with $O(n^8)$  states and a \twonfa\ $\mathcal{A}''$  with $O(n^8)$  states accepting the complement
  of the language accepted by $\mathcal{A}$.
\ecorollary

\bsection{Subexponential deterministic simulation}{s:deter}
In this section, we prove that each \twonfa\ with $n$ states can be simulated by
an equivalent \twdfa\ with  $O(n^{\log_2n+6})$ states, i.e.,
with a subexpontential, but still superpolynomial, number of states.
In the authors' knowledge this is the
first case of a model using nondeterminism and an unrestricted alphabet, having
a subexponential simulation by \twdfas.

This result generalizes a result proved for the unary case in~\cite{GMP03}.
Actually, even the proof is very similar: the new ``ingredient'' in
our version is the subroutine \reach\ presented in Section~\ref{s:reach}.
So we give a very short presentation, addressing the reader to~\cite{GMP03}
for further details.

Let $\mathcal{A}$ be a \twonfa\ with $n$ states in the form of Lemma~\ref{l:normalform}.
The \twdfa\ simulating $\mathcal{A}$ implements a recursive function, called
\reachable, based on the well known divide-and-conquere technique.

The function receives three parameters: two states $q$ and $p$ and an integer $t\geq 1$,
and returns a boolean. On these parameters,
$\reachable(q,p,t)$ returns true if and only if on the input $w$ under consideration
the automaton $\mathcal{A}$ has a sequence of at most $t$ segments {}from the state
$q$ to the state $p$.
Hence, according to Lemma~\ref{l:bound}, to decide whether or not $w$ is accepted by
$\mathcal{A}$, we call $\reachable(\qi,\qf,n-1)$. The pseudocode of the function
is given below.
We point out that is the base of the recursion, $t=1$, we have to verify the existence 
of a sequence of at most $1$ segment {}from $q$ to $p$. 
A sequence with $0$ segments is possible if and only if $q=p$. The existence of a sequence of just one
segment can be verified using the subroutine~\reach.

\bigskip

\begin{algorithm}[H]
  \caption{\emph{Recursive function $\reachable(q,p,t)$}}
\small
\DontPrintSemicolon
  \lIf{$t=1$}{\Return ($q=p$ or $\reach(q,p)$)\;\label{reachable_reach}
  }
  \Else{\label{reachable_>1}
      \ForEach{{\rm state} $r\in Q$\label{reachable_select}}{
        \If{$\reachable(q,r,\lceil t/2\rceil )$\label{reachable_lk/2}}{
            \lIf{$\reachable(r,p,\lceil t/2\rceil )$\label{reachable_rk/2}}{\Return\emph{true}\;}
        }
      }
    \Return\emph{false}\;\label{reachable_reject}
  }
\end{algorithm}

\bigskip
\noindent
We can prove the following:

\btheorem{t:deter}
  Each $n$-state \twonfa\ can be simulated by an equivalent \twdfa\ with
  $O(n^{\log_2n+6})$ states.
\etheorem
\bproof
  The implementation of \reachable\ and its complexity analysis are very
  close to those given in~\cite{GMP03} for the unary case.
  We just outline a rough estimation of the state upper bound.

  First, we suppose that the given \twonfa\ $\mathcal{A}$ is in the form given
  in Lemma~\ref{l:normalform}.
  The implementation of the function \reachable\ can be done using a constant height stack,
  as in~\cite{GMP03}, with the following differences:
  \begin{itemize}
  \item the height of the stack is $\lceil\log_2(n-1)\rceil$
  (instead of $\lceil\log_2(n+1)\rceil$),
  \item the subroutine \reach\ uses $4n-3$ state (the corresponding subroutine
  in~\cite{GMP03} was implemented with $n^2+3$ states).
  \end{itemize}
  This leads to an upper limit for the number of different stack configurations
  $4n(2n)^{\lceil\log_2(n-1)\rceil}$ (for the sake of simplicity, we deleted some negative terms,
  so it could be possible to give a slightly more accurate upper bound), which is
  bounded by $4n^{\lceil\log_2(n-1)\rceil+2}$.
  If the automaton $\mathcal{A}$ is not in form of Lemma~\ref{l:normalform}, we need to convert it
  using $3n$ states. Hence,
  by replacing $n$ by $3n$ in the above upper bound, we obtain the following rough upper estimation:
  \begin{eqnarray*}
  4(3n)^{\lceil\log_2(3n-1)\rceil+2} &\leq& 36\cdot n^2\cdot(3n)^{\lceil\log_2(3n-1)\rceil}\\
  &< &36\cdot n^2\cdot 9n^2 \cdot n^{\lceil\log_2(n-{1/3})+\log_23\rceil}\\
  &=&O(n^{\log_2(n)+c})
  \end{eqnarray*}
  We can observe that $c<6$.
\eproof

It is natural to wonder if the upper bound stated in Theorem~\ref{t:deter} is optimal.
We remind the reader that the best known lower bound for the number of the states of
\twdfas\ simulating $n$-state \twnfas\ is $O(n^2)$~\cite{Chr86}.
In the next section we will show that the optimality of the upper bound in
Theorem~\ref{t:deter} or any other superpolynomial state lower bound for the simulation
of \twonfas\ by \twdfas\ would imply the separation between deterministic and nondeterministic
logarithmic space, hence solving a longstanding open problem in structural complexity.

\bsection{Conditional and unambiguous simulations}{s:gap}

In this section we discuss how to reduce the language accepted by a given \twonfa\
to the \emph{graph accessibility problem} (\gap), i.e., the problem of deciding
whether a directed graph contains a path connecting two designated vertices.
This problem is well known to be complete for \nlogspace, the class of languages
accepted by $O(\log n)$ space bounded machines.
As a consequence of this reduction we will prove that the equality between the
classes $\logspace$ and $\nlogspace$ would imply a polynomial simulation of \twonfas\ by
\twdfas. Furthermore, we also prove that each \twonfa\ can be made unambiguous with
a polynomial increasing in the number of the states. An extension of this
reduction to the alternating case will be discussed in Section~\ref{s:alt}.

\medskip
Let us start to present our reduction. As for the results in Sections~\ref{s:compl}
and~\ref{s:deter}, it is obtained by combining a technique developed for the
unary case~\cite{GP11}, with the use of the subroutine~\reach\ presented
in Section~\ref{s:reach}.

{}From now on, we consider an $n$-state \twonfa\ $\mathcal{A}$ in the form of
Lemma~\ref{l:normalform}.
With each input string $w\in\Sigma^*$ we associate the directed graph $G(w)=(Q,E(w))$,
where $Q$ is the set of states of $\mathcal{A}$ and
\[
E(w)=\set{(p,q)\in Q\times Q\mid \reach(p,q) \mbox{ is true}},
\]
i.e., it is the set of pairs $(p,q)$ of states such that $\mathcal{A}$
on input $w$ has a segment {}from $p$ to $q$.

It should be clear that $w$ is accepted by $\mathcal{A}$
if and only if the graph $G(w)$ contains a path {}from vertex $\qi$,
the initial state, to vertex $\qf$, the accepting state.
Hence, this defines a reduction {}from the language accepted by
$\mathcal{A}$ to \gap.

As mentioned before, \gap\ is complete for \nlogspace\ under
logarithmic space reductions~\cite{Sav70}. Hence,
$\gap\in\logspace$ if and only if $\logspace=\nlogspace$.
This permit us to prove the following:%
\footnote{The result in Theorem~\ref{t:gap} is presented,
in a different context, also in~\cite{KP1X}.}

\btheorem{t:gap}
  If $\logspace=\nlogspace$ then each $n$-state \twonfa\ can be simulated
  by a \twdfa\ with a number of states polynomial in $n$.
\etheorem
\bproof
  Let us denote by $L$ be the language accepted by the given
  $n$-state \twonfa\ $\mathcal{A}$ and by \dgap\ a deterministic
  machine which solves \gap\ in logarithmic space. Such a machine
  exists, under the hypothesis $\logspace=\nlogspace$.

\begin{figure}[hbt]\centering
  \setlength{\unitlength}{0.275cm}
\begin{picture}(49,6)
\put(0,1){
  \begin{picture}(21,6)(0,0.5)

  \put(0,3){\vector(1,0){4}}
  \put(1,4){\makebox(0,0){\small$w$}}

  \put(4,1){\framebox(4,4){\small$G$}}

  \put(10,4){\makebox(0,0){\small$G(w)$}}
  \put(8,3){\vector(1,0){4}}

  \put(12,1){\framebox(4,4){\small$\dgap$}}

  \put(16.05,3.2){\vector(2, 1){3}}
  \put(19.2,4.7){\makebox(0,0)[l]{\small yes}}
  \put(16.05,2.8){\vector(2,-1){3}}
  \put(19.2,1.3){\makebox(0,0)[l]{\small no}}

  \put(3,0.5){\framebox(14,5){ }}

  \end{picture}
}
\put(28,0){
    \begin{picture}(21,6)(0,-0.5)

  \put(0,3){\vector(1,0){4}}
  \put(1,4){\makebox(0,0){\small$w$}}

  \put(4,1){\framebox(4,4){\small$G$}}

  \put(10,4){\makebox(0,0){\small$G(w)$}}
  \put(8,3){\vector(1,0){4}}

  \put(12,1){\framebox(4,4){\small$\ugap$}}

  \put(16.05,3.2){\vector(2, 1){3}}
  \put(19.2,4.7){\makebox(0,0)[l]{\small yes}}
  \put(16.05,2.8){\vector(2,-1){3}}
  \put(19.2,1.3){\makebox(0,0)[l]{\small no}}

  \put(3,-0.5){\framebox(14,6){ }}

  \put(11,0.75){\makebox(0,0)[t]{\small advice}}
  \put(11,0.75){\line(0,1){1.25}}
  \put(11,2){{\vector(1,0){1}}}
  \end{picture}
}
\end{picture}
\caption{The machines $M$ of Theorem~\ref{t:gap} (on the left) and $M_u$ of
Theorem~\ref{t:ugap} (on the right).}\label{f:gap}
\end{figure}
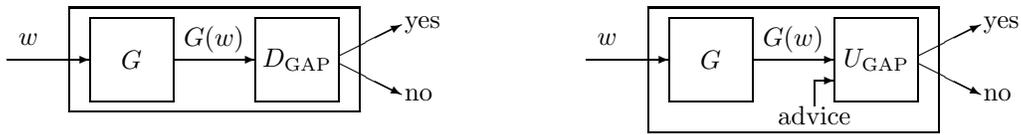

  We can build a machine $M$ deciding whether or not a string $w$ belongs
  to $L$ by composing the reduction $G$ above described with the machine
  \dgap\ (see Figure~\ref{f:gap}).

  The input of \dgap\ is the graph $G(w)$, more precisely its adjacency matrix,
  which can be encoded with $O(n^2)$ bits.
  Hence, \dgap\ uses  $O(\log n)$ space on its
  worktape. This space can be represented in a finite state control
  with a number of states polynomial in $n$. (Notice that since the automaton
  $\mathcal{A}$ is fixed, $n$ does not depend on the length of the input $w$.)

  On the other hand, the adjacency matrix of $G(w)$ cannot be stored
  in the finite  control of $M$, because it would require a number of
  states exponential in $n$. Hence, each time  \dgap\ needs to access
  one symbol {}from its input, such a bit is computed ``on the fly''.
  The bit represents a position in the adjacency matrix, corresponding to
  a pair $(p,q)$ of states of $\mathcal{A}$. Its value is~$1$ if and only if
  the automaton $\mathcal{A}$ has a segment {}from $p$ to $q$ on input $w$.
  Hence, it can be computed by calling $\reach(p,q)$, the subroutine
  presented in Section~\ref{s:reach}, which uses $4n-3$ states.
  The proof can completed as in~\cite[Lemma 4.1]{GP11}, where a similar
  result has been proved for the unary case.
\eproof
While the deterministic simulation in Theorem~\ref{t:gap} is under the condition
$\logspace=\nlogspace$, the next simulation by unambiguous machines
does not require any extra condition:

\btheorem{t:ugap}
  Each $n$-state \twonfa\ can be simulated by an \emph{unambiguous} \twonfa\ with a number of states
  polynomial in $n$.
\etheorem
\bproof
  In~\cite{RA00} it was proved for each nondeterministic machine $M$ working in logarithmic space
  there exists an equivalent unambiguous machine $M'$ still working in logarithmic space, which
  uses of a \emph{polynomial advice}~\cite{KL82}, i.e., of a family
  of strings $\set{\alpha(n)\mid n\geq 0}$ of polynomial length. The machine $M'$ receives,
  besides an input string $x$, the advice string corresponding to the length of $x$,
  i.e., the string $\alpha(|x|)$. Hence, there exists an unambiguous machine \ugap\ of
  this kind that, with an appropriate advice, solves \gap.

  Given a \twonfa\ $\mathcal{A}$ with $n$ states,
  by applying a construction similar to that in the proof of Theorem~\ref{t:gap},
  we can obtain an unambiguous two-way automaton $M_u$ with a number of states polynomial in $n$
  that recognize the same language as $\mathcal{A}$. Essentially, in the construction we have
  to replace the machine \dgap\ by the machine \ugap\ with the appropriate advice
  (see Figure~\ref{f:gap}).
  However, the advice depends only on the size of the input of \ugap, namely on the size
  of the graph $G(w)$, which, in turns, depends only on the number of states of~$\mathcal{A}$,
  not on the input string $w$. So, given the \twonfa\ $\mathcal{A}$, the advice is
  fixed, hence it can be encoded in the ``hardware'' of $M_u$.

  Finally, we observe that $M_u$ accesses its input tape only to
  compute the bits of the adjacency matrix of $G(w)$, i.e., when
  the subroutine \reach\ is running.
  This subroutine is deterministic. It starts and ends its computation with the head
  at left endmarker. Hence, when the machine \ugap\ takes nondeterministic
  decisions the head is scanning the left endmarker.
  This permit us to conclude that the unambiguous machine $M_u$
  is  a \twonfa.
\eproof

\bsection{The alternating case}{s:alt}

In this section we briefly discuss an extension of the technique used in Section~\ref{s:gap},
to the case of automata with~\emph{alternations}~\cite{CKS81}, recently considered
in~\cite{Kap09,Gef11}.

A \emph{two-way alternating automaton} (\twafa, for short) is
defined as a quintuple $\mathcal{A}=(Q,\Sigma,\delta,\qi,F)$,
exactly as a \twnfa. However, the set $Q$ is partitioned in two
sets $Q_{\exists}$ and $Q_{\forall}$, the sets of
\emph{existential} and \emph{universal} states, respectively.
The acceptance of an input string $w$ is witnessed by an
\emph{accepting computation tree} $T$.
The nodes of $T$ are labeled by configurations,
the root is labeled by the initial configuration, the leaves
are labeled by accepting configurations
(we can assume that accepting configurations are also halting).
Each node of $T$ labeled by a configuration $c$ whose state
is \emph{existential}  has exactly one son. The label
of this son is one of the possible successors of $c$.
Each node of~$T$ labeled by a configuration $c$ whose state
is \emph{universal} has one son for each possible successor of~$c$.
Notice that nondeterministic automata are just alternating automata
without universal states.

\medskip
Even for \twafas, we can restrict the use of the nondeterminism as we did for
2\nfas, considering \emph{outer-alternating finite automata}~(\twoafas). In these models,
each configuration scanning an ordinary input symbol can have at most
one successor, namely (existential or universal) nondeterministic
choices can be taken only when the head is scanning one of the
endmarkers.
Actually, we can further restrict the use of the nondeterminism only at the
left endmarker. In particular, using arguments similar to those of
Lemma~\ref{l:normalform}, we can prove the following result:\footnote{Notice a
small difference with respect to Lemma~\ref{l:normalform}. Stationary
moves are possible only at the left endmarker.
However, here they can be used also to reach states different than $\qf$.
Due to the presence of both
universal and existential states, the removal of these moves would require
an argument more complicated than the simple one used to prove
Lemma~\ref{l:normalform}. However, for our purposes, we do not need to
remove them.}

\blemma{l:altnormalform}
  For any $n$-state \twoafa~$\mathcal{A}=(Q,\Sigma,\delta,\qi,F)$
  there exists an equivalent \twoafa~$\mathcal{A}'$ with no more than $3n$ states
  that satisfies the following properties:
  \begin{enumerate}
    \item nondeterministic choices are taken only when the input head is scanning the left endmarker,
    \item there is a unique accepting state $\qf$ and this state is also halting,
    \item $\qf$ is reachable only at the left endmarker by stationary moves,
    \item stationary moves can occur only at the left endmarker.
  \end{enumerate}
\elemma
Now, we consider the \emph{alternating graph accessibility problem} (\agap, for short),
an alternating version of \gap.
The instance of the problem is an \emph{alternating} direct graph,
i.e., a graph $G=(V,E)$ with a partition of $V$ in two sets $V_{\exists}$ and $V_{\forall}$,
and two designated vertices $s$ and $t$. The question is if the predicate $apath(s,t)$ is true,
where, for $x,y\in V$, $apath(x,y)$ holds true if and only if:
\begin{itemize}
\item $x=y$, or
\item $x\in V_{\exists}$ and exists $z\in V$ with $(x,z)\in E$ such that $apath(z,y)$ is true, or
\item $x\in V_{\forall}$ and for all $z\in V$, $(x,z)\in E$ implies that $apath(z,y)$ is true.
\end{itemize}
This problem is known to be complete for the class $\ptime$, with respect to logarithmic space
reductions~\cite{Imm81}.

As is Section~\ref{s:gap}, we can reduce the language accepted by a given  \twoafa\
$\mathcal{A}$ to \agap, by associating with each input string $w$ the graph $G(w)=(Q,E(w))$,
where $(p,q)\in E(w)$ if and only if $\mathcal{A}$ has a computation segment {}from $p$ to $q$ on input $w$.
(The extension of the notion of computation segment to \twafa\ is obvious.)
Since the subroutine \reach\ presented in Section~\ref{s:reach} depends only on the transition function
of the given automaton~$\mathcal{A}$ and not on the acceptance condition,
we can use it to detect segments even in the case of outer \twafas.\footnote{Due to the possibility of stationary
moves on the left endmarker, the subroutine \reach\ needs just one small change (line~\ref{R_accept} in
the implementation presented in Appendix~\ref{a:reach}).
Given the two parameters $q'$ and $q''$, it preliminary checks if $q'=q''$ or if $q''$ is reachable
{}from $q'$ by a stationary move on the left endmarker. If the answer is positive, it immediately accepts,
otherwise, it works as we described.}

These permit us to prove the following result:

\btheorem{t:agap}
  If $\logspace=\ptime$ then each \twoafa\ can be simulated by a \twdfa\ with a polynomial
  number of states.
\etheorem
\bproof
  Since \agap\ is complete for \ptime, if $\logspace=\ptime$ then there exists a deterministic
  machine $\dagap$ that in logarithmic space solves $\agap$.
  Applying the same construction of Theorem~\ref{t:gap}, just replacing the machine
  \dgap\ by the machine \dagap, given an $n$-state \twoafa~$\mathcal{A}$, we can define
  an equivalent \twdfa\ with a number of states polynomial in $n$.
\eproof

In a similar way, we are able to prove the following:

\btheorem{t:nagap}
  If $\nlogspace=\ptime$ then each \twoafa\ can be simulated by a \twnfa\ with a polynomial
  number of states.
\etheorem
\bproof
  Under the hyphothesis $\nlogspace=\ptime$, there exists a nondeterministic
  machine \nagap\ that in logarithmic space solves $\agap$.
  By replacing in the previous proof the machine \dagap\ by the machine \nagap\
  we obtain a \twnfa\ with a number of states polynomial in $n$ equivalent to
  the given \twoafa.
\eproof

\bsection{Concluding remarks}{s:conclusion}
In this paper we generalized, in a unified framework, some
results previously obtained for unary \twnfas\ to machines with
arbitrary input alphabets, but making nondeterministic choices
only at the input tape endmarkers. Among others, we have shown
that any superpolynomial lower bound for the simulation of such
machines by standard \twdfas\ would imply
$\logspace\neq\nlogspace$\@. (Unlike in~\cite{BL77}, there are no
restrictions on the length of potential witness inputs\@.) In
Section~\ref{s:alt}, we also related the alternating version of
such machines to $\logspace\qeq\nlogspace\qeq\ptime$, the
classical computational complexity open problems.

Comparing our results with those obtained for other restricted
models of two-way automata we observe that:
\bitemize
  \item Actually, unary \twnfas\ can use only a restricted form of
    nondeterminism. In fact, we can restrict their nondeterminism
    to the endmarkers without increasing significantly their
    size~\cite{GMP03}\@. (A similar phenomenon has been observed by
    Chrobak in the case of unary \emph{one-way}
    automata~\cite{Chr86}\@.)
  \item In the general case, the possibility of reversing the input
    head movement at any input position does not seem so powerful as
    the possibility of making nondeterministic decisions at any input
    position. (Compare our polynomial upper bound for the
    complementation of \twonfas\ with the exponential lower bound for
    the complementation of sweeping \twnfas\ in~\cite{Kap06}\@.)
  \item However, in the deterministic case, the possibility of
    reversing the input head at any input position can make automata
    exponentially smaller than machines reversing the input head
    only at the endmarkers~\cite{Ber80,Mic81}\@.
\eitemize

It would be interesting to see if the results proved in this
paper could not be extended to a model using the nondeterminism
in a less restricted way than the one considered here.

\bibliographystyle{alpha}
\bibliography{biblio}

\begin{thebibliography}{DHRS97}

\bibitem[Ber80]{Ber80}
Piotr Berman.
\newblock A note on sweeping automata.
\newblock In Jaco de~Bakker and Jan van Leeuwen, editors, {\em Automata,
  Languages and Programming}, volume~85 of {\em Lecture Notes in Computer
  Science}, pages 91--97. Springer, 1980.

\bibitem[BL77]{BL77}
Piotr Berman and Andrei Lingas.
\newblock On the complexity of regular languages in terms of finite automata.
\newblock Technical Report 304, Polish Academy of Sciences, 1977.

\bibitem[Chr86]{Chr86}
Marek Chrobak.
\newblock Finite automata and unary languages.
\newblock {\em Theoretical Computer Science}, 47:149--158, 1986.
\newblock Errata: \emph{ibid}:302, 497-498.

\bibitem[CKS81]{CKS81}
Ashok~K. Chandra, Dexter Kozen, and Larry~J. Stockmeyer.
\newblock Alternation.
\newblock {\em J. ACM}, 28(1):114--133, 1981.

\bibitem[DHRS97]{DHRS97}
Pavol Duris, Juraj Hromkovi\v{c}, Jos{\'e} D.~P. Rolim, and Georg Schnitger.
\newblock Las vegas versus determinism for one-way communication complexity,
  finite automata, and polynomial-time computations.
\newblock In R{\"u}diger Reischuk and Michel Morvan, editors, {\em STACS},
  volume 1200 of {\em Lecture Notes in Computer Science}, pages 117--128.
  Springer, 1997.

\bibitem[Gef11]{Gef11}
Viliam Geffert.
\newblock An alternating hierarchy for finite automata.
\newblock In {\em Non-Classical Models of Automata and Applications (NCMA
  2011)}, pages 15--36, 2011.

\bibitem[GMP03]{GMP03}
Viliam Geffert, Carlo Mereghetti, and Giovanni Pighizzini.
\newblock Converting two-way nondeterministic unary automata into simpler
  automata.
\newblock {\em Theor. Comput. Sci.}, 295:189--203, 2003.

\bibitem[GMP07]{GMP07}
Viliam Geffert, Carlo Mereghetti, and Giovanni Pighizzini.
\newblock Complementing two-way finite automata.
\newblock {\em Inf. Comput.}, 205(8):1173--1187, 2007.

\bibitem[GP11]{GP11}
Viliam Geffert and Giovanni Pighizzini.
\newblock Two-way unary automata versus logarithmic space.
\newblock {\em Inf. Comput.}, 209(7):1016--1025, 2011.

\bibitem[HS03]{HS03}
Juraj Hromkovi\v{c} and Georg Schnitger.
\newblock Nondeterminism versus determinism for two-way finite automata:
  Generalizations of {S}ipser's separation.
\newblock In Jos C.~M. Baeten, Jan~Karel Lenstra, Joachim Parrow, and
  Gerhard~J. Woeginger, editors, {\em ICALP}, volume 2719 of {\em Lecture Notes
  in Computer Science}, pages 439--451. Springer, 2003.

\bibitem[HU79]{HU79}
John~E. Hopcroft and Jeffrey~D. Ullman.
\newblock {\em Introduction to Automata Theory, Languages and Computation}.
\newblock Addison-Wesley, 1979.

\bibitem[Imm81]{Imm81}
Neil Immerman.
\newblock Number of quantifiers is better than number of tape cells.
\newblock {\em Journal of Computer and System Sciences}, 22(3):384--406, 1981.

\bibitem[Kap06]{Kap06}
Christos~A. Kapoutsis.
\newblock Small sweeping 2{NFA}s are not closed under complement.
\newblock In Michele Bugliesi, Bart Preneel, Vladimiro Sassone, and Ingo
  Wegener, editors, {\em ICALP (1)}, volume 4051 of {\em Lecture Notes in
  Computer Science}, pages 144--156. Springer, 2006.

\bibitem[Kap09]{Kap09}
Christos~A. Kapoutsis.
\newblock Size complexity of two-way finite automata.
\newblock In Volker Diekert and Dirk Nowotka, editors, {\em Developments in
  Language Theory}, volume 5583 of {\em Lecture Notes in Computer Science},
  pages 47--66. Springer, 2009.

\bibitem[Kap11a]{Kap11b}
Christos~A. Kapoutsis.
\newblock Nondeterminism is essential in small 2{FA}s with few reversals.
\newblock In Luca Aceto, Monika Henzinger, and Jiri Sgall, editors, {\em ICALP
  (2)}, volume 6756 of {\em Lecture Notes in Computer Science}, pages 198--209.
  Springer, 2011.

\bibitem[Kap11b]{Kap11a}
Christos~A. Kapoutsis.
\newblock Two-way automata versus logarithmic space.
\newblock In Alexander~S. Kulikov and Nikolay~K. Vereshchagin, editors, {\em
  CSR}, volume 6651 of {\em Lecture Notes in Computer Science}, pages 359--372.
  Springer, 2011.

\bibitem[KL82]{KL82}
R.M. Karp and R.J. Lipton.
\newblock {Turing Machines that Take Advice}.
\newblock In E.~Engeler et~al, editor, {\em {Logic and Algorithmic}}.
  L'Enseignement Math{\'e}matique, Gen{\`e}ve, 1982.

\bibitem[KP11]{KP1X}
Christos~A. Kapoutsis and Giovanni Pighizzini.
\newblock Two-way automata characterizations of {L}/poly versus {NL}.
\newblock Submitted manuscript, 2011.

\bibitem[Mic81]{Mic81}
Silvio Micali.
\newblock Two-way deterministic finite automata are exponentially more succinct
  than sweeping automata.
\newblock {\em Information Processing Letters}, 12(2):103--105, 1981.

\bibitem[RA00]{RA00}
Klaus Reinhardt and Eric Allender.
\newblock Making nondeterminism unambiguous.
\newblock {\em SIAM Journal on Computing}, 29(4):1118--1131, 2000.

\bibitem[Sav70]{Sav70}
Walter~J. Savitch.
\newblock Relationships between nondeterministic and deterministic tape
  complexities.
\newblock {\em J. Comput. Syst. Sci.}, 4(2):177--192, 1970.

\bibitem[Sip80a]{Sip80h}
Michael Sipser.
\newblock Halting space-bounded computations.
\newblock {\em Theor. Comput. Sci.}, 10:335--338, 1980.

\bibitem[Sip80b]{Sip80}
Michael Sipser.
\newblock Lower bounds on the size of sweeping automata.
\newblock {\em J. Comput. Syst. Sci.}, 21(2):195--202, 1980.

\bibitem[SS78]{SS78}
William~J. Sakoda and Michael Sipser.
\newblock Nondeterminism and the size of two way finite automata.
\newblock In {\em STOC}, pages 275--286. ACM, 1978.

\bibitem[Sze94]{Sze94}
Andrzej Szepietowski.
\newblock {\em Turing Machines with Sublogarithmic Space}, volume 843 of {\em
  Lecture Notes in Computer Science}.
\newblock Springer, 1994.

\end{thebibliography}

\newpage
\appendix
\bsection{The subroutine $\reach$}{a:reach}

Let $\mathcal{A}$ be a \twonfa\ in the form of Lemma~\ref{l:normalform}.
Fixed $q',q''\in Q$, we first describe a \twdfa\ $\mathcal{A}'$ which halts on every input and accept a word $w$
if and only if $\mathcal{A}$ on input $w$ has a computation segment {}from $q'$ to $q''$.
Subsequently, we will discuss how to modify the construction to use $q'$ and $q''$ as parameters.

First, we examine some trivial cases:
\begin{itemize}
\item If $q'=q''$ then $\mathcal{A}'$ can immediately accept.
\item If $q'\neq q''$ and $\delta(q',\vdash)=\emptyset$ then $\mathcal{A}'$ can immediately reject.
\item If $q'\neq q''$ and $q''=\qf$ then if $(\qf,0)\in\delta(q',\vdash)$ then $\mathcal{A}'$ immediately
  accepts, otherwise it immediately rejects.
\end{itemize}
Now, we consider the remaining cases. Since the construction we present is obtained by
modifying that given in~\cite{GMP07} for making \twdfas\ halting,
the presentation will be given along the same lines and by keeping, as much as possible,
the same notations.

First, let us fix a linear order on the state set of the original automaton.
As usual, the symbols ``$<$'' and~``$>$'' denote the ordering relation.
We do not need to further consider the final state $\qf$ and stationary moves.
In fact, the only possible stationary moves are those reaching $\qf$ on the left endmarker,
$\qf$ cannot be reached in any other input position and it is halting (see Lemma~\ref{l:normalform}).
All these possibilities have been already considered in the case of segments ending in $\qf$.

Our implementation of the depth-first search examines each configuration $(q,i)$ in two modes:
\begin{itemize}
\item\emph{Mode 1:} Examination of the ``left'' predecessors of $(q,i)$, namely
    configurations $(p,i-1)$ such that $(q,+1)\in\delta(p,a)$,
\item\emph{Mode 2:} Examination of the ``right'' predecessors, namely configurations $(p,i+1)$ such that
    $(q,-1)\in\delta(p,a)$.
\end{itemize}
For each $q\in Q$ and both modes, we introduce a starting and a finishing state.
So our machine will use the following set of states:
\[Q'=\set{\ql,\, \qhl,\, \qr,\, \qhr: q\in Q-\set{\qf}}\cup\set{q_Y}.\]
These states are interpreted as follows:
\begin{itemize}
    \item[\rm $\ql$] Starting state for the Mode~1: examination of left predecessors for the configuration $(q,i)$.
      Left predecessors will be examined one after another, according to the linear order induced by the relation~``$<$''.
      To inspect the content of the input square $i-1$, the simulator (if it is in the state~$\ql$)
      has its input head one position to the left of the actual position of the original
      machine~$\mathcal{A}$ in configuration $(q,i)$.
    \item[\rm $\qhl$] Finishing state for the Mode~1. All the
      left predecessors of $(q,i)$ have been examined, but we
      still have to examine the right predecessors of $(q,i)$.
      In the state~$\qhl$, the input head of the simulator is in the actual position, i.e., the position~$i$.
    \item[\rm $\qr$] Starting state for the Mode~2, examination of right predecessors for $(q,i)$,
      when the left predecessors have been finished.
      The right predecessors will also be examined in the linear order induced by~``$<$''.
      In the state~$\qr$, the simulator has its input head one position to the right of the actual position
      of the configuration $(q,i)$, to inspect the symbol~$a$ in the input square $i+1$.
    \item[\rm $\qhr$] Finishing state for the Mode~2.
      Both the left and the right predecessors of $(q,i)$ have been
      examined. In the state~$\qhr$, the input head of the simulator is in
      the actual position, i.e., the position~$i$.
    \item[\rm $q_Y$] Halting accepting state.
\end{itemize}
Let us now describe the transition function
$\delta':Q'\times(\Sigma\cup\set{\vdash,\dashv})\rightarrow Q'\times\set{-1,0,+1}$
of~$\mathcal{A}'$ implementing this strategy.
For each (type of) state $r\in Q'$ and each symbol $a\in\Sigma\cup\set{\vdash,\dashv}$,
we first display a procedure that assigns a value of $\delta'(r,a)\in Q'\times\set{-1,0,+1}$ to the transition table
and, after that, we present an explanation for this procedure.
Note that the simulator will use stationary moves.
The reader should also keep in mind that the procedures displayed below are
\emph{not} executed by the machine~$\mathcal{A}'$
but, rather, they are used to fill in the entries in the transition table for~$\mathcal{A}'$.
In the description, we will suppose that the input string is $w=w_1w_2\ldots w_n$, with
$w_i\in\Sigma$, for $i=1,\ldots,n$.

\bigskip

\NoCaptionOfAlgo
\begin{algorithm}[H]
  \caption{\emph{Transition $\delta'($\ql$,a)$}}
\small
\DontPrintSemicolon
\lIf{$a = \dashv$}{$\delta'(\ql,a):=$ undefined\;\label{ql_dashv}}
\uElseIf{$a = \vdash$\label{ql_vdash}}{
        \lIf{$(q,+1)\in\delta(q',\vdash)$}{$\delta'(\ql,a):=(q_Y,0)$\;\label{ql_testvdash}}
        \lElse{$\delta'(\ql,a):= (\qhl,+1)$\;\label{ql_transfervdash}}}
\lElseIf{{\rm there is no} $p\in Q:\delta(p,a) = \set{(q,+1)}$}{{$\delta'(\ql,a):= (\qhl,+1)$\;\label{ql_empty}}
        \Else{
            $\tilde{q}:=\min\set{p\in Q:\delta(p,a)= \set{(q,+1)}}$\label{ql_select}\;
            $\delta'(\ql,a):= (\Ql{\tilde{q}},-1)$\label{ql_transfer}\;
            }
}
\end{algorithm}
\bigskip
\noindent
This part presents several differences with the corresponding one in~\cite{GMP07}.
Recall that $\mathcal{A}'$ gets to the state~$\ql$
when, for some~$i$, it starts the examination of the left predecessors of the configuration $(q,i)$,
that is, of configurations $(p,i-1)$ such that $(q,+1)\in\delta(p,w_{i-1})$.
By definition of~$\ql$, $\mathcal{A}'$~has its input head already at the position $i-1$.
The procedure considers four cases:
\begin{itemize}
\item $a=\dashv$ (line~\ref{ql_dashv}): actually this case is unreachabale, it is given just for
  completeness, to fill in all entries in the transition table for~$\delta'$.
\item $a=\vdash$ ({}from line~\ref{ql_testvdash}): in this case the original machine $\mathcal{A}$ is
  scanning the first input cell, i.e., that containing $w_1$.
  There are two possibilities depending on the first parameter~$q'$.
  If {}from the state $q'$ on the left endmarker it is possible to move right in the state $q$,
  i.e., $(q,+1)\in\delta(q',\vdash)$, then
  the backward simulation of the segment is completed. Hence, the machine halts and
  accept (line~\ref{ql_testvdash}).
  Otherwise, since we are interested just in \emph{one} segment, we ignore the left predecessors
  of $(q,1)$, since they correspond to moves {}from the left endmarker, i.e., to the starting of
  segments others than the one we are interested in. The machine
  immediately enters in the ending state of Mode 1 $\qhl$, moving right the input head, to reach
  the real position in the simulated machine (line~\ref{ql_transfervdash}).
\item There are no left predecessors (line~\ref{ql_empty}): the machine ends Mode~1 (notice that,
  since we already considered $a=\vdash$, in this case and in the next one the transitions we have to
  consider are deterministic).
\item In the remaining case (lines~\ref{ql_select}--\ref{ql_transfer}),
  there exists at least one left predecessor of configuration $(q,i)$.
  We select the first one in the linear order and start to examine this configuration
  with the same method. To this aim, we switch the state to~$\Ql{\tilde{q}}$,
  and move the head one position to the left of $i-1$.
\end{itemize}

\bigskip

\NoCaptionOfAlgo
\begin{algorithm}[H]
  \caption{\emph{Transition $\delta'($\qhl$,a)$}}
\small
\DontPrintSemicolon
\lIf{$a\neq\dashv$}{$\delta'(\qhl,a):=(\qr,+1)$\;\label{qhl_transfer}}
\lElse{$\delta'(\qhl,a):=(\qhr,0)$\;\label{qhl_dashv}}
\end{algorithm}
\bigskip
\noindent
This case is exactly as in~\cite{GMP07}.
In this state, the examination of the left predecessors of
$(q,i)$ has been completed. Hence, the search continues with the
examination of the right predecessors in the Mode~2 (line~\ref{qhl_transfer}), by
switching to the state~$\qr$ and moving the head to the
position~$i+1$.
If the input head is already on the right endmarker, i.e.,
$a=\,\dashv$, then the configuration $(q,i)$ does not have any
right predecessors (line~\ref{qhl_dashv}). Hence, by switching to~$\qhr$, we
finish the Mode~2 immediately, as if all right predecessors had
been searched.

\bigskip

\NoCaptionOfAlgo
\begin{algorithm}[H]
  \caption{\emph{Transition $\delta'($\qr$,a)$}}
\small
\DontPrintSemicolon
\lIf{$a = \vdash$}{$\delta'(\qr,a):=$ undefined\;\label{qr_vdash}}
    \lElseIf{{\rm there is no} $p\in Q:\delta(p,a) = \set{(q,-1)}$}{$\delta'(\qr,a):= (\qhr,-1)$\;\label{qr_empty}}
    \Else{
        $\tilde{q}:=\min\set{p\in Q:\delta(p,a)= \set{(q,-1)}}$\;\label{qr_select}
        $\delta'(\qr,a):= (\Ql{\tilde{q}},-1)$\;\label{qr_transfer}
}
\end{algorithm}
\bigskip
\noindent
Even this case is as in~\cite{GMP07}.
In the state $\qr$, $\mathcal{A}'$~starts to examine right predecessors of $(q,i)$,
i.e., configurations $(p,i+1)$ such that $\delta(p,w_{i+1})=(q,-1)$.
$\mathcal{A}'$~has its head already at the position $i+1$.
We observe that in a right predecessor of a configuration, the head
cannot scan the left endmarker. Hence, all transitions {}from right
predecessors are deterministic.
There are three main cases:
\begin{itemize}
\item $a=\vdash$ (line~\ref{qr_vdash}): unreachabale case given  for  completeness.
\item There are no right predecessors (line~\ref{qr_empty}): we finish the Mode~2 immediately, which completes the search for $(q,i)$.
\item Otherwise (lines~\ref{qr_select}--\ref{qr_transfer}) we select $(\tilde{q},i+1)$,
the first right predecessor of $(q,i)$, and start to examine it with the same method.
(Among others, the left predecessors of $(\tilde{q},i+1)$ are going to be examined.)
To this aim, we switch to $\Ql{\tilde{q}}$,
and move the head one position to the left of $i+1$.
\end{itemize}

\bigskip

\NoCaptionOfAlgo
\begin{algorithm}[H]
  \caption{\emph{Transition $\delta'($\qhr$,a)$}}
\small
\DontPrintSemicolon
\lIf{$a=\vdash$}{$\delta'(\qhr,a):=$ undefined\;\label{qhr_reject}}
\lElseIf{$\delta(q,a) = \emptyset$}{$\delta'(\qhr,a):=$ undefined\;\label{qhr_undefined}}
    \Else{
        $(r,d)$ := unique element of $\delta(q,a)$\label{qhr_successor}\;
        \uIf{{\rm there is no} $p\in Q: p>q$ {\rm and} $\delta(p,a) = \set{(r,d)}$\label{qhr_empty}}{
            \lIf{$d=+1$}{$\delta'(\qhr,a):=(\Qhl{r},+1)$\;\label{qhr_emptyleft}}
            \lElse{$\delta'(\qhr,a):=(\Qhr{r},-1)$\;\label{qhr_emptyright}}}
        \Else{
            $\tilde{q}:= \min\set{p\in Q: p>q \mbox{ and } \delta(p,a) = \set{(r,d)}}$\;\label{qhr_select}
            $\delta'(\qhr,a):=(\Ql{\tilde{q}},-1)$\;\label{qhr_transfer}
     }
}
\end{algorithm}
\bigskip
\noindent
This state concludes the examination of the configuration $(q,i)$,
and all configurations in the subtree rooted at $(q,i)$.
The machine~$\mathcal{A}'$ has its head at the position~$i$, the actual position
of the head of the simulated machine $\mathcal{A}$.
There are three main cases:
\begin{itemize}
\item The head is scanning the left endmarker (line~\ref{qhr_reject}):
  since in the backward simulation we ignore the transitions {}from the left endmarker
  (see line~\ref{ql_transfervdash} in the definition of $\delta'(\ql,a)$),
  the only reachable configuration of the form $(q,0)$ that can be reached
  is $(q'',0)$, namely, the configuration {}from which the backward simulation started.
  If this happens, it means that we have examined all the tree rooted at $(q'',0)$, never accepting.
  Hence, a computation segments {}from $(q',0)$ to $(q'',0)$
  on input $w$ does not exists and $\mathcal{A}'$ have to reject.
  This can be done by leaving the transition undefined: the computation
  stops in a state different {}from the accepting state $q_Y$.
\item The configuration does not have any successor (line~\ref{qhr_undefined}):
  a such configuration is never reached in the backward search. This case is
  included only for completeness.
\item In the remaining case ({}from line~\ref{qhr_successor}), since the head is scanning
  a position different {}from the left endmarker, the configuration $(q,i)$ has as unique
  successor $(r,j)$, which can obtained using the transition function of $\mathcal{A}$
  (line~\ref{qhr_successor}), $\delta(q,a)=\set{(i,d)}$ and $j=i+d$.
  Depending on the value of $d$, we have to consider
  either left predecessors ($d=+1$) or right predecessors ($d=-1$) of $(r,j)$
  (we call them ``$d$-predecessors'', for short).
  First we try to find a state $p$ greater than $q$ such that $(p,i)$ is a $d$-predecessor.
  If it does not exists then $(q,i)$ is the last $d$-predecessor of $(r,j)$.
  Hence (depending on $d$), we complete Mode 1 or Mode 2 for $(r,j)$
  (lines~\ref{qhr_emptyleft}--\ref{qhr_emptyright}).
  Otherwise, we start to examine in Mode~1, with the same method, the next $d$-predecessor
  of $(r,j)$ (lines~\ref{qhr_select}--\ref{qhr_transfer}).
\end{itemize}
To complete the description of the automaton $\mathcal{A}'$, we have to specify that the initial
state is $\Qhl{q''}$.
In fact, we are interested in a segment {}from $q'$ to $q''$ on input $w$.
By definition, such a segment should end in the configuration $(q'',0)$.
This configuration does not have left predecessors.
This permits us to state the depth-first search {}from the state $\Qhl{q''}$,
with the head on the left endmarker.
If there exists the segment we are looking for, then the computation ends, as explained before,
in the state $q_Y$ (line~\ref{ql_testvdash}).
Otherwise, it  ends in some different state (line~\ref{qhr_reject}) after traversing the whole
subtree rooted at $(q'',0)$.

\medskip
We observe that the automaton $\mathcal{A}'$ so described has $4n-3$ states
and it depends on the two states $q'$ and $q''$ we have fixed at the beginning.
What happens if we change those states, i.e., if we consider them as parameters?

The only parts of the construction that depend on $q'$ and $q''$ are the choice of the initial state
and the transitions {}from the endmarkers and states $\ql$ (lines~\ref{ql_testvdash}--\ref{ql_transfervdash}).
We can modify $\mathcal{A'}$ by keeping, for each possible value  of the parameter $q'$,
a table with the values of $\delta'(\ql,\vdash)$.
During the execution $\mathcal{A'}$ will consult the appropriate table
depending on the value of the parameter. All the other transitions of $\mathcal{A'}$
are unchanged. Furthermore, we can initialize $\mathcal{A'}$ in a suitable state,
depending on the parameter $q''$.

Hence, using $4n-3$ states (not including the space used to store the two parameters $q',q''$),
we can implement $\reach(q',q'')$ with the following procedure
(which also considers the trivial cases and which never modify the two parameters):

\bigskip

\NoCaptionOfAlgo
\begin{algorithm}[H]
  \caption{$\reach(q',q'')$}
\small
\DontPrintSemicolon
\lIf{$q'=q''$}{accept\;}\label{R_accept}
\lElseIf{$q'\neq q''$ and $\delta(q',\vdash)=\emptyset$}{reject\;}
\uElseIf{$q'\neq q''$ and $q''=q_{acc}$}{
    \lIf{$(q_{acc},0)\in\delta(q',\vdash)$}{accept\;}
    \lElse{reject}\;}
\lElse{run $\mathcal{A'}$ {}from the state $\Qhl{q''}$\;}
\end{algorithm}
\bigskip
\noindent
By summarizing, we have shown that given an $n$-state \twonfa\ $\mathcal{A}$ in the form of
Lemma~\ref{l:normalform},
using a finite state control consisting of $4n-3$ states we can decide whether or not for an input
string $w$ and two states $q'$ and $q''$ the automaton $\mathcal{A}$ has a computation segment {}from
$q'$ to $q''$ on $w$, hence proving Lemma~\ref{l:lemma}.

\bsection{The subroutines $\nreach$ and $\treach$}{a:treach}

Given a state $q''$, $\nreach(q'')$ should return a nondeterministic chosen state $q'$ such that
there exists a computation segment {}from $q'$ to $q''$ on the given input string.
If the subroutine is not able to find such a segment, due to the fact that such a state $q'$
does not exist or due to wrong nondeterministic choices, then the subroutine halts in the state~\qques,
aborting the entire computation.

The implementation can be given as a modification of that of \reach. In particular,
$\nreach(q'')$ uses the backward search described in Section~\ref{a:reach}, starting {}from
the configuration $c_f=(q'',0)$. When a configuration $c$ with the input head scanning the first
input symbol is reached, i.e., $c=(q,1)$, then the subroutine nondeterministically selects
one between these two actions:
\begin{itemize}
\item Output a nondeterministically chosen state $q'$ such that $(q,+1)\in\delta(q',\vdash)$.
  This means that $(q',0)$ is a left predecessor of $(q,1)$ and so there is a segment {}from $q'$ to $q''$.
  (If such a state $q'$ does not exists, the computation is aborted in~\qques.)
\item Ignore the left predecessors of $(q,1)$ and  continue the backward simulation. Hence,
  the machine try to detect a longer segment.
\end{itemize}
These actions can be implemented by replacing the selection on lines~\ref{ql_testvdash}
and~\ref{ql_transfervdash}, with a nondeterministic choice between these two possibilities:
\begin{itemize}
\item $q''$:= a nondeterministically choosen state $q'$ such that $(q,+1)\in\delta(q',\vdash)$
  (if such a state does not exist the entire computation is aborted in~\qques).\\
  $\delta'(\ql,a):=$ undefined\\
  This stop the computation by producing the output in the parameter, as a side effect.
\item $\delta'(\ql,a):= (\qhl,+1)$
\end{itemize}
The computation can also traverse the whole subtree rooted at $c_f=(q'',0)$ without reaching the
starting configuration of a segment. This can happen  either because a segment ending in $q''$
does not exist or because a wrong sequence of nondeterministic choices. In this case we have
to halt in the state~\qques, aborting the entire computation. In the implementation of
$\reach$ this case was managed in line~\ref{qhr_reject}. We need simply to change the
assignment into $\delta'(\qhr,a):=\qques$.

\medskip
This implementation of \nreach\ uses the same set of states of \reach, with the exception of
$q_Y$, plus the ``global'' state~\qques. Hence it can implemented with $4n-3$ states.
We should also notice that it can produce, as a side effect, a modification of the parameter.

The nondeteministic subroutine $\treach$ is implemented calling $t$
times $\nreach$ as shown in the following high-level code:\footnote{The (redundant) assignment on
 line~\ref{simulation_call} is to
emphasize that $\nreach(q'')$ leaves its output in the parameter.}

\bigskip

\NoCaptionOfAlgo
\begin{algorithm}[H]
  \caption{$\treach(q'',t)$}
\small
\DontPrintSemicolon
$\tilde{t} := t$\;
\While{$\tilde{t}>0$\label{simulation_while}}{
    $q'':=\nreach(q'')$\tcp*{possible side effect of \nreach:\ abort in $\qques$}\label{simulation_call}
    $\tilde{t}:=\tilde{t} - 1$\;
}
\lIf{$q''\neq q_0$}{halt in~\qques\;}
\end{algorithm}

\end{document}